\documentclass[letterpaper, 10pt, conference]{ieeeconf} 

\let\labelindent\relax

 \makeatletter
 \let\NAT@parse\undefined
 \makeatother
 
\usepackage[numbers,sort&compress]{natbib}
\usepackage{float,setspace, graphicx,enumitem, psfrag, amssymb, amsmath, color, epstopdf, epsfig}

\usepackage{graphics} 
\usepackage{amsmath} 
\usepackage{amssymb}  
\usepackage{amsthm}
\usepackage{bm}
\usepackage{acronym}
\usepackage{paralist}
\usepackage{float}
\usepackage{color}
\usepackage{epstopdf}
\usepackage{multicol}
\usepackage{tikz}
\usepackage{hyperref}

\usepackage[mathscr]{euscript}
\usepackage{xcolor}
\newfont{\smalll}{cmr8}
\def\IR{\mathbb{R}}
\def\ID{\mathbb{D}}

\def\span{\mathrm{span}}

\def\IC{\hbox{C\hskip-
.5em\raise.5ex\hbox{$\scriptscriptstyle\mid$}}\ }
\def\Ic{\hbox{\smalll C\hskip-
.5em\raise.3ex\hbox{$\scriptscriptstyle\mid$}}\ }

\def\T={\buildrel {\scriptscriptstyle\triangle} \over =}

\def\sqr#1#2{{\vcenter{\vbox{\hrule height.#2pt\hbox{\vrule
width.#2pt height#1pt \kern#1pt\vrule width.#2pt}\hrule
height.#2pt}}}}

\def\inf{\mathop{\rm inf}}
\def\block-diag{\mathop{\rm block{\scriptstyle -}diag}}

\def\pmbb#1{\setbox0=\hbox{#1}\raise 0.5ex\box0}

\def\norm#1{\|#1\|}

\newcommand{\bequ}{\begin{eqnarray}}
\newcommand{\eequ}{\end{eqnarray}}
\newcommand{\mT}{^\mathrm{T}}

\newcommand{\rom}{\mathrm}

\newcommand {\beq}      {\begin{equation}}
\newcommand {\eeq}      {\end{equation}}

\newcommand{\Proj}{{\mathrm{Proj}}}

\def\IR{{\mathbb R}}
\def\IC{{\mathbb C}}

\renewcommand\theenumi{\roman{enumi}}
\renewcommand\theenumii{\alph{enumii}}
\renewcommand\theenumiii{\arabic{enumiii}}
\renewcommand\theenumiv{\Alph{enumiv}}

\def\sc{s_{\rm c}}

\def\sc{s_{\rm c}}

\newcommand{\vertiii}[1]{{\left\vert\kern-0.25ex\left\vert\kern-0.25ex\left\vert #1 
		\right\vert\kern-0.25ex\right\vert\kern-0.25ex\right\vert}}

\newtheorem{rem}{Remark}
\newtheorem{assump}{Assumption}
\newtheorem{theo}{Theorem}
\newtheorem{defin}{Definition}

\definecolor{tRed}{RGB}{250,40,0}
 
\definecolor{tBlue}{RGB}{0,100,250}

\makeatletter
\hypersetup{colorlinks=true}
\AtBeginDocument{\@ifpackageloaded{hyperref}
  {\def\@linkcolor{blue}
  \def\@anchorcolor{red}
  \def\@citecolor{red}
  \def\@filecolor{red}
  \def\@urlcolor{red}
  \def\@menucolor{red}
  \def\@pagecolor{red}
\begingroup
  \@makeother\`%
  \@makeother\=%
  \edef\x{%
    \edef\noexpand\x{%
      \endgroup
      \noexpand\toks@{%
        \catcode 96=\noexpand\the\catcode`\noexpand\`\relax
        \catcode 61=\noexpand\the\catcode`\noexpand\=\relax
      }%
    }%
    \noexpand\x
  }%
\x
\@makeother\`
\@makeother\=
}{}}
\makeatother

\IEEEoverridecommandlockouts

\begin{document}

\title{\LARGE \bf Safety-Critical Adaptive Control with \\ Nonlinear Reference Model Systems}
\author{Ehsan Arabi \and Kunal Garg \and Dimitra Panagou
\thanks{\hspace{.18cm}Ehsan Arabi is a Postdoctoral Research Fellow of the Department of Aerospace Engineering at the University of Michigan, Ann Arbor, MI 48109, USA (email: {\tt earabi@umich.edu}).
\newline\indent Kunal Garg is a Ph.D. Candidate of the Department of Aerospace Engineering at the University of Michigan,  Ann Arbor, MI 48109, USA (email: {\tt kgarg@umich.edu}).
\newline\indent Dimitra Panagou is an Assistant Professor of the Department
of Aerospace Engineering and the Director of the Distributed Aerospace Systems and Control Laboratory at University of Michigan, Ann Arbor, MI 48109, USA (email: {\tt dpanagou@umich.edu}).
}
\thanks{The authors would like to acknowledge the support of the Air Force Office of Scientific
Research under award number FA9550-17-1-0284.}}

\maketitle
\begin{abstract}
	In this paper, a model reference adaptive control architecture is proposed for uncertain nonlinear systems to achieve prescribed performance guarantees. 
	Specifically, a general nonlinear reference model system is considered that captures an ideal and safe system behavior. An adaptive control architecture is then proposed to suppress the effects of system uncertainties without any prior knowledge of their magnitude and rate upper bounds. 	More importantly, the proposed control architecture enforces the system state trajectories to evolve within a user-specified prescribed distance from the reference system trajectories, satisfying the safety constraints. This eliminates the ad-hoc tuning process for the adaptation rate that is conventionally required in model reference adaptive control to ensure safety. The efficacy of the proposed control architecture is also demonstrated through an illustrative numerical example.
\end{abstract}
\vspace{-.04cm}

\section{Introduction} \label{sec1} \vspace{-.03cm}
Adaptive control systems are control algorithms that mitigate the effects of system uncertainties and exogenous disturbances.
However, one of the limiting factors of these control systems is their lack of verifiable system performance.
Model reference adaptive control generally consists of a reference model system and a control architecture along with an update law. In the design of the update law, the choice of the adaptation rate plays a crucial role in the overall system performance and in how much the system trajectories deviate from the reference model trajectories (i.e., from the ideal system behavior). 
As a result, the ad-hoc tuning process of the adaptation rate that is essential for safety-critical applications for keeping the system trajectories within the safe set, usually relies heavily on excessive vehicle testing and hence is time-consuming and costly.  

To address this challenge, model reference adaptive control algorithms are proposed to achieve strict performance guarantees in  \cite{yucelen2014adaptive,muse2011method,arabi2017set,arabi2018var,l2018constrained,l2018barrier}. Similar to most of model reference adaptive control literature, the reference model systems in these studies follow linear dynamics. 
However, nonlinear reference systems are preferable for several practical applications, especially for those involving guidance and control of highly-maneuverable aircraft, guided projectiles, and space launch vehicles.
Notable contributions to the adaptive control literature using nonlinear reference systems are documented by the authors of \cite{wagg2003adaptive,yucelen2015adaptive,wang2012l1,kara2017model,kawaguchi2007passivity,peter2012adaptive,scarritt2008nonlinear}.
In particular, \cite{wagg2003adaptive} proposes an adaptive control algorithm for scalar nonlinear systems based on a nonlinear reference system, while \cite{yucelen2015adaptive} extends this result for a general class of uncertain nonlinear systems. 
In \cite{wang2012l1}, the $\mathcal{L}_1$ adaptive control method is used with nonlinear reference models. 
A sliding mode control design is proposed in \cite{kara2017model}   using the state-dependent Riccati equation.
Furthermore, applications of adaptive control with nonlinear reference model systems in active steering systems, tail-controlled missiles, and satellite attitude control are studied respectively in  \cite{kawaguchi2007passivity}, \cite{peter2012adaptive} and \cite{scarritt2008nonlinear}.
Yet, the aforementioned approaches do not establish any strict performance guarantees on the system trajectories, and they may violate the safety requirements specially during the transient time. Therefore, in safety-critical applications, a control designer either requires \textit{a-priori} and almost complete knowledge of upper and lower bounds on the system uncertainties, or need to perform an ad-hoc tuning process for rendering the closed-loop system trajectories within the safe set (see \cite{arabi2017set} and references therein for more details).

Our contribution is to present and analyze a new model reference adaptive control architecture based on nonlinear reference models, with strict performance guarantees. 
Specifically, the proposed control architecture suppresses effects of system uncertainties independently of their magnitude and rate upper bounds, and enforces the system state trajectories to evolve within a user-specified prescribed distance from the reference system states, satisfying the system safety constraints. 
For the case when the reference trajectories are available prior to implementation,  a time-varying performance bound is imposed on the system error vector. When the available information is limited to a set of reference system trajectories, a constant performance bound is imposed, which is characterized based on the minimum distance of the reference set and the boundary of the safe set. This result can be viewed as a generalization of the results in \cite{arabi2017set,arabi2018var} where a set-theoretic model reference adaptive control is proposed for linear dynamical systems with linear reference models. In fact, the presented results in this paper reduce to the control algorithms in \cite{arabi2017set,arabi2018var} for a special case (see Remark \ref{set_theory}). 
An illustrative numerical example is also provided to demonstrate the efficacy of the proposed architectures. 

\section{Mathematical Preliminaries} \label{sec2}

We begin with the notation used in this paper. 
$\IR$, $\IR^n$, and $\IR^{n \times m}$ respectively denote the set of real numbers,
the set of $n \times 1$ real column vectors, 
and the set of $n \times m$ real matrices;
$\IR_+$ (resp., $\overline{\mathbb R}_+$) and $\IR_+^{n \times n}$ denote the set of positive real numbers (resp., non-negative reals) and the set of $n \times n$ positive-definite  real matrices;
$\ID^{n \times n}$ denotes the set of $n \times n$ diagonal matrices,
$\textrm{bd}(S)$ denotes the boundary of the set $S\subset\mathbb R^n$,
and ``$\triangleq$''  denotes equality by definition. 
In addition, we use $(\cdot)\mT$ to denote the transpose operator,
$(\cdot)^{-1}$ to denote the inverse operator, 
 $\mathrm{det}(\cdot)$ to denote the determinant operator, 
 $\norm{\cdot}$ to denote the Euclidean norm,
$\| \cdot \|_\rom{F}$ to denote the Frobenius norm,
$\lambda_{\min}(A)$ (resp., $\lambda_{\max}(A)$) to denote the minimum (resp., maximum) eigenvalue of the square matrix $A$,
$\textrm{dist}(A,B) = \inf_{x\in A, y\in B}\|x-y\|$ to denote the distance between the sets $A, B \subset \IR^n$,
and $\textrm{dist}(x,B) = \inf_{y\in B}\|x-y\|$ to denote the distance of $x \in \IR^n$ from the set $B \subset \IR^n$.
The gradient of a continuously differentiable function $f:\mathbb R^n\rightarrow\mathbb R^m$, evaluated at $x\in \mathbb R^n$ is denoted as $\nabla f(x) \triangleq \frac{\partial f}{\partial x}(x)$.

Next, we introduce the definition of the projection operator. Let $\Omega$ be a convex hypercube in $\IR^n$ defined as $\Omega = \left\{\theta\in\IR^n : (\theta^\rom{min}_i \leq \theta_i \leq \theta^\rom{max}_i )_{i=1,2,\cdots ,n}\right\}$,  where $(\theta^\rom{min}_i,$ $\theta^\rom{max}_i)$ denote the minimum and maximum bounds for the $i\rom{th}$ component of the $n$-dimensional parameter vector $\theta$.
Furthermore, let $\Omega_\nu$ be the second hypercube defined as $\Omega_\nu = \left\{\theta\in\IR^n : (\theta^\rom{min}_i + \nu \leq \theta_i \leq \theta^\rom{max}_i - \nu)_{i=1,2,\cdots ,n}\right\}$, where $\Omega_\nu \subset \Omega$ for a sufficiently small positive constant $\nu$. 

{{\defin [\cite{lavretsky2012robust, pomet1992adaptive}] \label{proj}  
For $y \in \IR^n$, the projection operator $\Proj:\IR^n \times \IR^n \rightarrow \IR^n$ is defined (componentwise) as $\rom{Proj}(\theta,y) \triangleq \big({[\theta^\rom{max}_i - \theta_i]}/{\nu}\big)y_i$ when $\theta_i > \theta^\rom{max}_i - \nu$ and $y_i > 0$, $\rom{Proj}(\theta,y) \triangleq \big({[\theta_i - \theta^\rom{min}_i]}/{\nu}\big)y_i$ when $\theta_i < \theta^\rom{min}_i + \nu$ and $ y_i < 0$, and $\rom{Proj}(\theta,y) \triangleq y_i$ otherwise.
}}

\vspace{0.1cm}
It follows from Definition \ref{proj} that $\big(\theta-\theta^{*}\big)^\rom{T}\big(\mathrm{Proj}\left(\theta,y\right)-y\big)\leq 0$, $\theta^{*} \in \Omega_\nu$, where this inequality can be readily generalized to matrices using $\rom{Proj}_\rom{m}(\Theta, Y) = \bigl(\rom{Proj}($ $\rom{col}_{1}(\Theta), \rom{col}_{1}(Y)), \ldots, \rom{Proj}(\rom{col}_{m}(\Theta),\rom{col}_{m}(Y))\bigl)$
with $\Theta\in\IR^{n \times m}$, $Y\in\IR^{n \times m}$, and $\mathrm{col}_{i}(\cdot)$ denoting $i\rom{th}$ column operator.

	
	
	
	
	


\section{Problem Formulation} \label{sec3}
In this paper, we consider the class of uncertain nonlinear dynamical systems of the form
\bequ
\dot{x}(t) = F(x(t)) + G u(t) + D \delta(t,x(t)), \quad x(0) = x_{0}, \label{sys}
\eequ
where $x \in \IR^{n}$ is the system state vector, 
$F: \IR^{n} \rightarrow \IR^{n}$ is a known system vector field with $F(0)=0$, 
$G \in\IR^{n \times m}$ is an unknown control input matrix,
$u(t) \in \IR^m$ is the control input, 
$D \in\IR^{n \times m}$ is a known matrix, and
$\delta : \overline{\IR}_+ \times \IR^{n} \rightarrow \IR^m$ denotes system uncertainties. Let $S_s \subset \IR^n$ denote the safe set of system states such that $x(t) \in S_s$ ensures safety. Consider the nonlinear reference model dynamics capturing an ideal (and safe) system behavior given by 
\bequ
\dot{x}_\rom{r}(t) &=& F_\rom{r}(x_\rom{r}(t), c(t)), \quad x_\rom{r}(0)=x_{\rom{r}0}, \label{refsys}
\eequ
where $x_\rom{r} \in S_r \subset S_s$ is the reference system state vector, 
$c(\cdot)$ is a bounded command signal, and $F_\rom{r}: \IR^{n} \times \IR \rightarrow \IR^{n}$ is the reference system vector field. 
The control objective is to design an adaptive control signal $u(\cdot)$ for the uncertain nonlinear dynamical system in \eqref{sys} to suppress the effects of system uncertainties such that the system state $x(\cdot)$ tracks the reference system state $x_\rom{r}(\cdot)$ while maintaining safety, i.e. $x(t) \in S_s, \forall t\geq 0$.

\begin{figure}[t!] \center \vspace{0.0cm} \epsfig{file=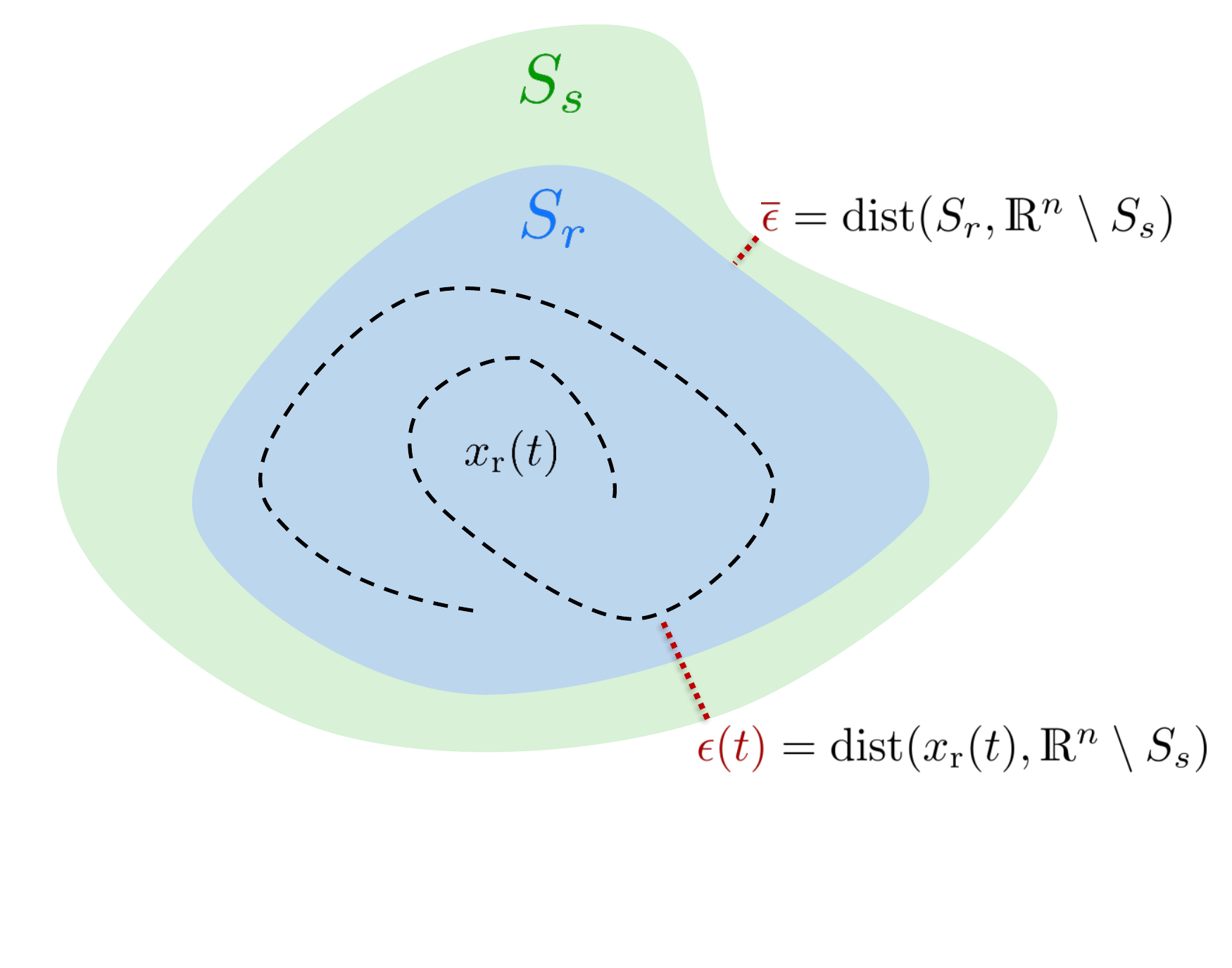, scale=.40} \vspace{-1.2cm}
	\caption{Graphical representation of the sets $S_s$ and $S_r$, and the performance bounds $\epsilon(t)$  and $\overline{\epsilon}$.} \vspace{-.3cm}
	\label{fig0} \end{figure}

Define the error vector between the system state trajectories and the reference system trajectories as $e(t) \triangleq x(t) - x_\rom{r}(t)$. 
If $e(t) \in \mathcal{D}_t$ where $\mathcal{D}_t \triangleq \{e : \norm{e} < \epsilon(t) \}$ with the time-varying performance bound $\epsilon(t)\triangleq \textrm{dist}(x_r(t),\mathbb R^n\setminus S_s) \in \IR_+$, then $x(t) \in S_s$, i.e., the trajectories of the uncertain dynamical system remain within the safe set $S_s$ (see Figure \ref{fig0}). In other words, if the control architecture limits the maximum deviation of the system state trajectories from the reference system by the performance bound $\epsilon(t)$, that is $\|e(t)\|\leq \epsilon(t)$ for all $t\geq 0$, then safety is guaranteed (see Remark \ref{TI} for case with a constant performance bound $\bar\epsilon$). This is a challenging task since the calculated upper bound on the system error signal in standard adaptive control designs is generally conservative, and depends on the upper bounds of system uncertainties \cite{arabi2017set}.
We now introduce a standard assumption on system uncertainty parameterization \cite{narendra2012stable,ioannou2012robust,lavretsky2012robust}. 

{{\assump \label{as_1}
The system uncertainty given by \eqref{sys} is parameterized as
\bequ
\delta(t,x(t))&=& W_\rom{p}\mT (t) \sigma_\rom{p}(x(t)), \label{eq0}
\eequ
where $W_\rom{p}(t) \in \IR^{s \times m}$  is a bounded time-varying unknown weight matrix with a bounded time rate of change (i.e., $\| W_\rom{p}(t) \|_\rom{F} \le w_\rom{p}$ and $\| \dot{W}_\rom{p}(t) \|_\rom{F} \leq {{w}_\rom{pd}}$ for some unknown $w_\rom{p}, {{w}_\rom{pd}} \in \IR_+$) and $\sigma_\rom{p} : \IR^{n} \rightarrow \IR^s$ is a known basis function of the form $\sigma_\rom{p}(x)=[\sigma_{\rom{p}1}(x), \sigma_{\rom{p}2}(x), \ . \ . \ . \ ,\sigma_{\rom{p}s}(x)]\mT$. 
}}

{{\assump \label{as_0}
The unknown control input matrix $G$ in \eqref{sys} is parameterized as
\bequ
G &=& D \Lambda , \label{eq1}
\eequ
where $\Lambda \in \IR_+^{m \times m} \cap  \ID^{m \times m}$ is a bounded unknown control effectiveness matrix. 
}}

%

{{\rem
The presented results in this paper can be readily extended to the class of nonlinear systems with state-dependent control matrix, i.e., $G = G(x(t))$ in \eqref{sys}, where  $G(x(t)) = D \Lambda  H(x(t))$ and $\rom{det}(H(x(t)) \neq 0$ hold \cite{volyanskyy2009new,yucelen2015adaptive}. }}

Using \eqref{sys} and \eqref{refsys} along with Assumptions \ref{as_1} and \ref{as_0}, the system error dynamics can be written as \vspace{0cm}
\bequ
\dot{e}(t) &\hspace{-.25cm}=\hspace{-.25cm}& F(x(t))-F_\rom{r}(x_\rom{r}(t), c(t)) + D \Lambda\big( u(t)\nonumber \\ &\hspace{-.25cm}\hspace{-.25cm}&  + \Lambda^{-1} W_\rom{p}\mT (t) \sigma_\rom{p}(x(t))\big), \quad  e(0) = e_0,   \label{eq2}
\eequ
with $e_0 \triangleq x_{0}-x_{\rom{r}0}$. In the absence of system uncertainties (i.e., $\delta(t,x(t)) \equiv 0$ and $\Lambda = I_{m\times m}$), one can write \eqref{eq2} as
\bequ
\hspace{-.35cm}\dot{e}(t) &\hspace{-.25cm}=\hspace{-.25cm}& F(x(t))-F_\rom{r}(x_\rom{r}(t), c(t))  + D u_\rom{n}, \label{eq3}
\eequ
where $u_\rom{n} = u_\rom{n}(x(t),x_\rom{r}(t),c(t)) \in \IR^m$ is a nominal control law. 

{{\assump  \label{AssumNom}
In the absence of system uncertainties (i.e., $\delta(t,x(t)) \equiv 0$ and $\Lambda = I_{m\times m}$), there exist a nominal control law $u_\rom{n} = u_\rom{n}(x(t),x_\rom{r}(t),c(t)) \in \IR^m$ such that the origin of the system error dynamics in \eqref{eq3} is exponentially stable with a continuously differentiable positive definite function $V: \mathcal{D}_t \rightarrow \IR$, $ \mathcal{D}_t \subset S_s$ satisfying
\bequ 
k_1 \norm{e(t)}^2 \leq V(e(t)) \leq  k_2 \norm{e(t)}^2, \label{Vbound} 
\eequ
$\forall e(t) \in \mathcal{D}_t$ and its time derivative satisfying
\bequ
\dot{V}(e(t)) &=& \nabla V\mT(e(t)) \ g(e(t)) \leq -M(e(t)),  \label{eq4}\\
k_0 \norm{e(t)}^2 &\leq& M(e(t)), \label{Wbound}
\eequ
where $\nabla V(e(t)) = {\partial V(e(t))}/{\partial e(t)}$, $g(e(t))=F(x(t))-F_\rom{r}(x_\rom{r}(t), c(t))  + D u_\rom{n}(x(t),x_\rom{r}(t),c(t))$ and $k_0, k_1$ and $k_2$ are positive constants. 
}}

{{\rem
The nominal control input $u_n$ satisfying Assumption \ref{AssumNom} can be found by various methods for some special classes of system: if \eqref{sys} and \eqref{refsys} are linear, then the LQR control input satisfies this assumption (see Remark \ref{set_theory}); if the said systems are polynomial, then sum-of-squares (SOS) techniques can be used to find the nominal controller (see \cite{zheng2009sum,peet2009exponentially} and references therein); or, a QP based method \cite{ames2016control} can be used to compute a control input for a larger class of nonlinear, control affine systems. 
}}

{{\rem \label{rem1}
Let $h(t,e(t)) \triangleq k_1\epsilon^2(t)- V(e(t))$. If $h(t,e(t)) >0, \forall t\geq0$, then $e(t) \in \mathcal{D}_t, \forall t\geq0$.
}}

{{\assump  \label{Assumepsidot}
The time derivative of $\epsilon(t)$ exists, and satisfies the condition $|\dot \epsilon| \leq \frac{\alpha_1}{2}\epsilon$ if $\dot \epsilon<0$, where $\alpha_1 = \frac{k_0}{k_2}$, for all $t\geq 0$.
}}

Note that Assumption \ref{Assumepsidot} limits the rate of change in the performance bound $\epsilon(t)$ only when this bound is decreasing. More importantly, since this bound is characterized based on the safe set $S_s$ and the reference system trajectories $x_\rom{r}(t)$ where $x_\rom{r}(t) \in S_s, \forall t\geq0$, one can always find a performance bound $\epsilon(t)$ such that it satisfies Assumption \ref{Assumepsidot}. 

\section{Adaptive Control Architecture with Performance Guarantees} \label{proposed}
In this section, we design and analyze an adaptive control architecture for enforcing a performance bound on the system error vector, limiting the deviation of system trajectories from the reference model trajectories. To this end, we rewrite \eqref{eq2}  as
\bequ
\dot{e}(t) &\hspace{-.25cm}=\hspace{-.25cm}& F(x(t))-F_\rom{r}(x_\rom{r}(t), c(t))  + D u_\rom{n}(x(t),x_\rom{r}(t),c(t)) \nonumber \eequ \bequ &\hspace{-.25cm}\hspace{-.25cm}& + D \Lambda\big( u(t) + \Lambda^{-1} W_\rom{p}\mT (t) \sigma_\rom{p}(x(t))  \nonumber \\ 
&\hspace{-.25cm}\hspace{-.25cm}& - \Lambda^{-1} u_\rom{n}(x(t),x_\rom{r}(t),c(t))\big), \quad  e(0) = e_0.   \label{eq5}
\eequ
Defining $W(t) = [\Lambda^{-1} W_\rom{p}\mT (t), -\Lambda^{-1}]\mT \in \IR^{(s+m)\times m}$ and $\sigma(x(t),x_\rom{r}(t),c(t)) = [\sigma_\rom{p}\mT(x(t)), u_\rom{n}\mT(x(t),x_\rom{r}(t),c(t))]\mT  \in \IR^{(s+m)}$, \eqref{eq5} can be expressed as
\bequ
\dot{e}(t)  = g(e(t)) + D \Lambda\big( u(t) +  W\mT (t) \sigma(\cdot)\big). 
\label{eq6}
\eequ
Note that $\| W(t) \|_\rom{F} \le w$ and $\| \dot{W}(t) \|_\rom{F} \leq {w_\rom{d}}$ automatically holds, for some unknown $w, {w_\rom{d}} \in \IR_+$ as a direct consequence of Assumption \ref{as_1}.
Motivated by the structure of the system error dynamics in \eqref{eq6}, let the adaptive control law be 
\bequ
u(t) &=& - \hat{W}\mT (t) \sigma(\cdot), \label{uad}
\eequ
where $\hat{W}\in  \IR^{(s+m)\times m}$ is an estimate of the unknown weight matrix $W$ satisfying the parameter adjustment mechanism 
\bequ
\dot{\hat{W}} (t) &\hspace{-.25cm}=\hspace{-.25cm}& \gamma \rom{Proj_m}\Big(\hat{W}(t), \frac{h(t,e)+V(e)}{h^2(t,e)}\sigma(\cdot) \nabla V\mT(e) D\Big), \nonumber \\
 {\hat{W}} (0) &\hspace{-.25cm}=\hspace{-.25cm}& {\hat{W}}_0,  \label{Wdot}
\eequ
with $\gamma \in \IR_+$ being a constant adaptation rate and $\hat{W}_\rom{max}$ being the projection operator bound.  

For the next theorem presenting the main result of this paper, we now write the system error dynamics and the weight estimation error dynamics respectively as
\bequ
\dot{e}(t)&\hspace{-.2cm}=\hspace{-.2cm}& g(e(t)) - D\Lambda \tilde{W}\mT (t)\sigma(\cdot)\big), \quad e(0)=e_0,   \label{err1} \\
\dot{\tilde{W}}(t)&\hspace{-.2cm}=\hspace{-.2cm}&\gamma \rom{Proj_m} \Big(\hat{W}(t), \frac{h(t,e)+V(e)}{h^2(t,e)}\sigma(\cdot) \nabla V\mT(e) D\Big)  \hspace{.4cm} \nonumber \\ &\hspace{-.2cm}\hspace{-.2cm}&- \dot{W}(t), \quad \tilde{W}(0)=\tilde{W}_0, \label{err2}
\eequ
where $\tilde{W} (t)\triangleq \hat{W}(t)- W(t)$ is the weight estimation error.

{{\theo \label{theo1} 
Consider the uncertain nonlinear dynamical system given by (\ref{sys}) subject to Assumptions \ref{as_1}-\ref{AssumNom},
the nonlinear reference model given by (\ref{refsys}) capturing an ideal system behavior,
and the feedback control law given by (\ref{uad}) along with (\ref{Wdot}). If $e_0 \in \mathcal{D}_t$, then the closed-loop dynamical system trajectories given by (\ref{err1}) and (\ref{err2}) are bounded, and $e(t) \in \mathcal{D}_t \  \forall t\geq 0$, i.e., the system state vector $x(t)$ remains within the safe set $S_s$ for all times.
}}

Due to page limitations, the proof of the above theorem will be reported elsewhere. Here, we only provide a sketch of the proof. 
Specifically, consider the energy function $\Psi:\mathcal{D}_t \times \IR^{(s+m)\times m}$ $\rightarrow \overline{\IR}_+$ given by 
\bequ
\Psi(e,\tilde{W}) = \frac{V(e)}{h(t,e)} + \frac{\gamma^{-1}}{2}\rom{tr}\bigl[(\tilde{W}(t)\Lambda^{1/2})\mT(\tilde{W}(t)\Lambda^{1/2})\bigr]. \hspace{-.5cm}\nonumber \\ \label{psi}
\eequ
The time derivative of \eqref{psi} along the closed-loop system trajectories (\ref{err1}) and (\ref{err2}) can be written as \vspace{-.2cm}

\bequ
\dot{\Psi}\bigl(e,\tilde{W}\bigl) 
&\hspace{-.3cm}\leq\hspace{-.3cm}& 
-\alpha_1 \Psi\bigl(e,\tilde{W}\bigl)  + \alpha_2-\alpha_1\Big(\frac{V(e)}{h(t,e)}\Big)^2 
  \nonumber \\ &\hspace{-.3cm}\hspace{-.3cm}& -2k_1\epsilon(t)\dot \epsilon(t)\frac{V(e)}{h^2(t,e)},  \label{psid4}
\eequ
where $\alpha_1 \triangleq {k_0}/{k_2} \in \IR_+$ and $\alpha_2 \triangleq d+ \alpha_1 \gamma^{-1} \tilde{w} \norm{\Lambda}/2  \in \IR_+$. 
If $\dot \epsilon(t)\leq 0$ and $V(e)<{2k_1\epsilon(t)|\dot\epsilon(t)|}/{\alpha_1}$ hold, it follows from definition of $h(t,e)$ and Assumption \ref{Assumepsidot} that 
\bequ
    h(t,e) >k_1\epsilon^2(t)-\frac{2k_1\epsilon(t)|\dot\epsilon(t)|}{\alpha_1}>0.
\eequ
Thus, the closed-loop system trajectories given by \eqref{err1} and \eqref{err2} are bounded and $e(t) \in \mathcal{D}_t$. 
If the above conditions for $\dot \epsilon(t)$ and $V(e)$ do not hold, one can write
\bequ
\dot{\Psi}\bigl(e,\tilde{W}\bigl)
&\hspace{-.3cm}\leq\hspace{-.3cm}& 
-\alpha_1 \Psi\bigl(e,\tilde{W}\bigl)  + \alpha_2\label{psid5}.
\eequ
It now follows that the energy function $\Psi\bigl(e,\tilde{W}\bigl)$ is upper bounded by $\Psi\bigl(e,\tilde{W}\bigl) \leq \Psi_{\max}$, where $\Psi_{\max} \triangleq \max\{\Psi_0,\alpha_2/\alpha_1\} \in \IR_+$, $\Psi_0 \triangleq {\Psi}(e(0),\tilde{W}(0)) \in \IR_+$, resulting in boundedness of the closed-loop system trajectories given by \eqref{err1} and \eqref{err2}. Furthermore, using \eqref{psi} one can write $\frac{V(e)}{h(t,e)} \leq \Psi_{\max}$. 
Hence, per Remark \ref{rem1}, $h(t,e(t))>0$, or equivalently, $e(t)\in \mathcal D_t$ for all $t\geq 0$, i.e., the system state $x(t)$ remains within the safe set $S_s$ at all times.

{{
\rem \label{set_theory}
Considering a linear system dynamics and a linear reference dynamics respectively as
\bequ
\dot{x}(t) &=& A x(t) + B \Lambda (u(t) + \delta(t,x(t))), \\ 
\dot{x}_\rom{r}(t) &=& A_\rom{r} x_\rom{r}(t) + B_\rom{r} c(t),
\eequ
Assumption \ref{AssumNom} is equivalent to the existence of the control gains $K_1\in \IR^{m\times n}$ and $K_2 \in \IR^{m\times n_c}$ known as matching conditions such that $A_\rom{r} = A-BK_1$ and $B_\rom{r}=BK_2$ hold \cite{lavretsky2012robust,nguyen2018model}.
In this special case, define the weighted Euclidean norm of system error as $||e(t)||_P = \sqrt{e\mT(t) P e(t)}$ where $P \in \IR_+^{n\times n}$ is a solution to the Lyapunov equation  $0 = A_\rom{r}\mT P + P A_\rom{r} + R$, $R \in \IR_+^{n\times n}$. By choosing $V(e(t)) = ||e(t)||^2_P$, the update law in \eqref{Wdot} reduces to
\bequ
\dot{\hat{W}} (t) \ &\hspace{-.3cm}=\hspace{-.3cm}& \ 2 \gamma \rom{Proj_m}\Big(\hat{W}(t),  \sigma(\cdot)\frac{ k_1 \epsilon^2(t) \ e\mT(t) P B}{(k_1 \epsilon^2(t) - ||e(t)||^2_P)^2} \Big), \ \ \ \ \ 
\eequ
with ${\hat{W}} (0)={\hat{W}}_0,$ which is in the same form as the proposed update law in \cite{arabi2017set,arabi2018var} for the set-theoretic model reference adaptive control using generalized restricted
potential functions (see (5) and (6) of \cite{arabi2018var}).
}}

{{
\rem \label{TI}
If it is desired to work with constant performance bound instead of the time-varying performance bound discussed in this section,  one can define $\bar \epsilon \triangleq \inf_{t\geq 0}\epsilon(t)$ and consider the time-invariant set $\bar{\mathcal D} \triangleq \{e(t)\; |\; \|e(t)\|\leq \bar \epsilon\}$ instead of $\mathcal D_t$.
In addition, consider the case when only a set of reference system trajectories is available for control design instead of the reference system trajectories prior to implementation. In this case, $x_\rom{r}(t)\in S_r$ for all $t\geq 0$, for some $S_r\subset S_s$; hence, one can define  a time-invariant performance bound as $\bar\epsilon \triangleq \textrm{dist}(S_r,\mathbb R^n\setminus S_s) \in \IR_+$ with the set $\bar{\mathcal D}$ defined similarly as above (see Figure \ref{fig0}).
}}

\section{Illustrative Numerical Example} \label{sim}
In this section, we present a numerical example to demonstrate the efficacy of the proposed control architecture. 
Specifically, we consider the uncertain dynamical system given by 
$\dot{x}(t) = F(x(t)) + D \Lambda u(t) + D \delta(t,x(t)), \quad x(0) = x_{0},$
with
$
F(x(t)) = \begin{bmatrix} 
x_2(t) \\
-x_1(t)-x_1(t) x_2(t)+x_2^2(t)  
\end{bmatrix}, 
D = \begin{bmatrix} 
0\\
1 
\end{bmatrix},$
where $x(t) = [x_1(t), x_2(t)]\mT$  $\in \IR^2$ denotes the system state vector. 
In addition, the unknown system uncertainty has the form 
$\delta(t,x(t)) =  0.3 \sin(0.1 t) x_1(t) + 0.3 \cos(0.3 t) x_1(t) x_2(t) + x_1(t) x_2^2(t)$.
We next consider the nonlinear reference model representing the forced Van der Pol oscillator \cite{yucelen2015adaptive,haddad2008nonlinear} given by
\bequ
\begin{bmatrix}
\dot{x}_\rom{r1}(t)\\
\dot{x}_\rom{r2}(t)
\end{bmatrix} = \begin{bmatrix} 
x_\rom{r2}(t) \\
-x_\rom{r1}(t)+\mu x_\rom{r2}(t) (1-x^2_\rom{r1}(t))+ c(t)
\end{bmatrix}, 
\eequ
with $x_\rom{r}(0)=x_\rom{r0}$ and $c(t) =  1.2 \sin(t)$.
In the absence of system uncertainties (i.e., $\delta(t,x(t)) \equiv 0$ and $\Lambda = 1$), the nominal controller
$
u_\rom{n}(x(t),x_\rom{r}(t),c(t)) = -l_1 e_1(t) - l_2 e_2(t) + x_1(t) x_2(t)    - x_2^2(t) + c(t) + \mu x_\rom{r2}(t) (1-x^2_\rom{r1}(t)),$
with $l_1, l_2 \in \IR_+$, satisfies Assumption \ref{AssumNom}, resulting in the exponentially stable error dynamics $\dot{e}(t) = A_e e(t), \ e(0) = e_0$ with
$A_e =
\begin{bmatrix}
0  & 1 \\
-(1+l_1)  & -l_2
\end{bmatrix}$,
and the Lyapunov function $V(e(t)) = e\mT(t) P e(t)$ where $P \in \IR_+^{2\times2}$ is a solution to the Lyapunov equation  $0 = A_e\mT P + P A_e + R$, $R \in \IR_+^{2\times2}$.

\begin{figure}[b!] \center \vspace{.0cm}
	\includegraphics[width=8cm]{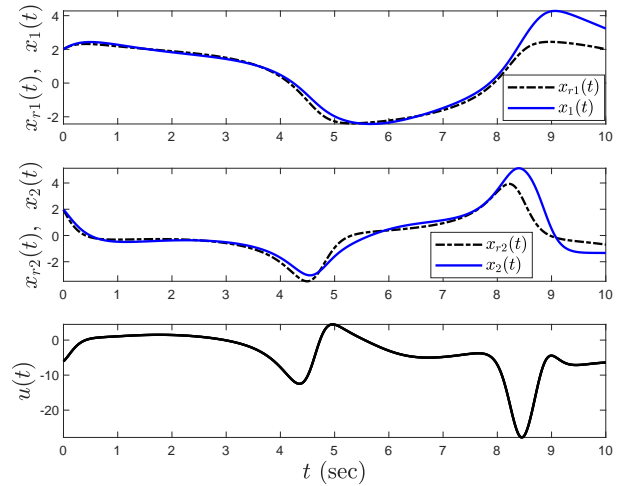} \vspace{-0.27cm}
	\caption{System performance with the nominal controller.} \vspace{-.0cm}
	\label{fig1} \end{figure}

\begin{figure}[t!] \center \vspace{-.00cm}
	\includegraphics[clip,trim=.2cm 0cm .8cm .7cm, width=8.5cm]{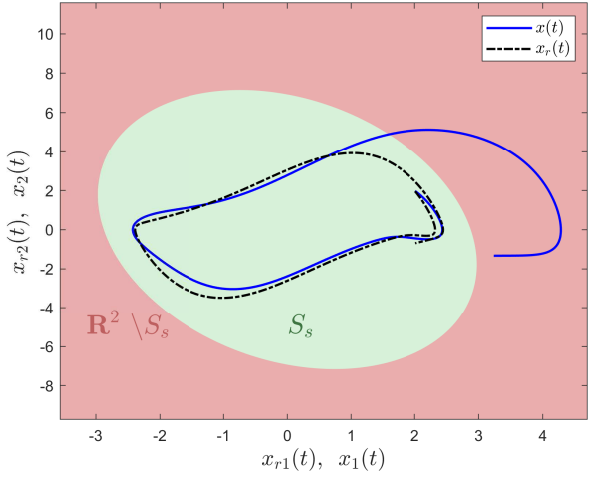} \vspace{-0.3cm}
	\caption{System state phase portrait with the nominal controller.} \vspace{-.15cm}
	\label{fig2} \end{figure}

We set the initial conditions to $x_0 = x_\rom{r0}=[2,2]\mT$, $\mu=1$, $l_1=l_2=3$, and the control effectiveness to $\Lambda = 0.75$. 
In addition, the safe set for system trajectories is chosen as $S_s = \{x(t): x\mT(t) P x(t) < 3.2 \}$. 
Figures \ref{fig1} and \ref{fig2} present the performance of the nominal controller in the presence of system uncertainties. It is evident that the nominal controller is not capable of keeping the system state trajectories within the safe set $S_s$.
We now consider two cases to illustrate how the proposed results with constant and time-varying performance bounds are used.

\begin{figure}[t!] \center \vspace{.0cm}
	\includegraphics[clip,trim=.2cm 0 .8cm .7cm, width=8.5cm]{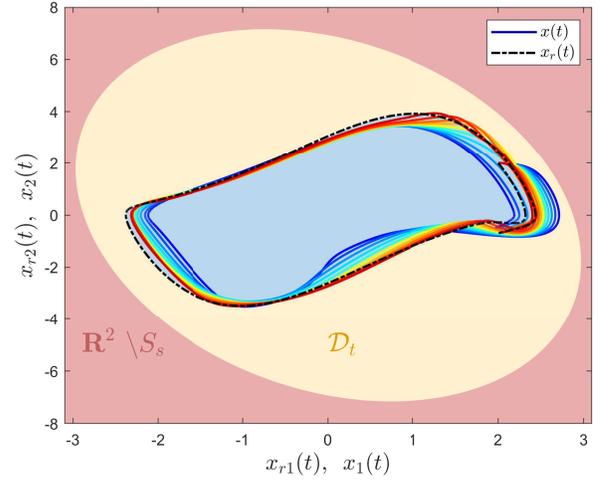} \vspace{-0.3cm}
	\caption{System state phase portrait with the proposed adaptive control architecture  with time-varying performance bound for $\gamma \in [0.05, 5]$ (blue to red).} \vspace{-.0cm}
	\label{fig8tv} \end{figure}

\begin{figure}[t!] \center \vspace{-.3cm}
	\includegraphics[clip,trim=.2cm 0 .8cm .7cm, width=8.5cm]{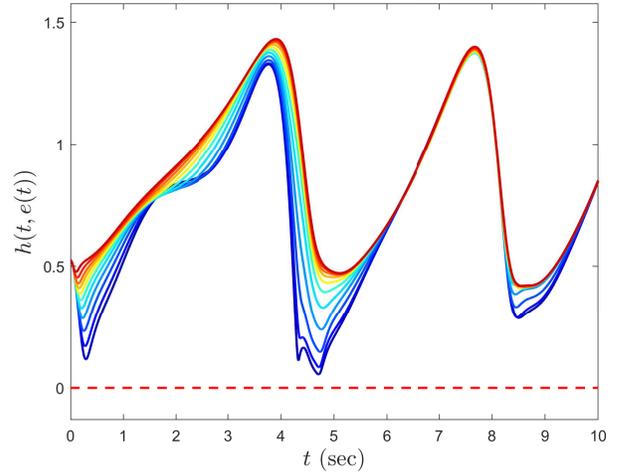} \vspace{-0.27cm}
	\caption{The evolution of $h(e(t))$ with the proposed adaptive control architecture  with time-varying performance bound for $\gamma \in [0.05, 5]$ (blue to red).} \vspace{-.0cm}
	\label{fig9tv} \end{figure}

First, we consider that the reference system trajectories $x_\rom{r}(t)$ is known prior to the implementation. In this case, the safety margin can be characterized as a time-varying set $\mathcal{D}_t$, and the time-varying performance bound $\epsilon(t)$ is selected based on the distance of the reference system trajectories from the safe set $S_s$ defined above at each time instance.
We apply the adaptive control signal in \eqref{uad} with the proposed update law in \eqref{Wdot} with different values of the adaptation rate $\gamma \in [0.05, 5]$. 
Figure \ref{fig8tv} shows that although with lower adaptation rates system state gets closer to the boundaries of the safe set $S_s$, they never leave this set. This is also clear from  Figure \ref{fig9tv} where $h(t,e(t))$ is always positive resulting in $e(t) \in \mathcal{D}_t \ \forall t\geq0$.
This shows that the obtained performance guarantee is independent of the selection of the adaptation rate $\gamma$ as expected.

\begin{figure}[t!] \center \vspace{-.3cm}
	\includegraphics[clip,trim=.2cm 0 .8cm .7cm, width=8.5cm]{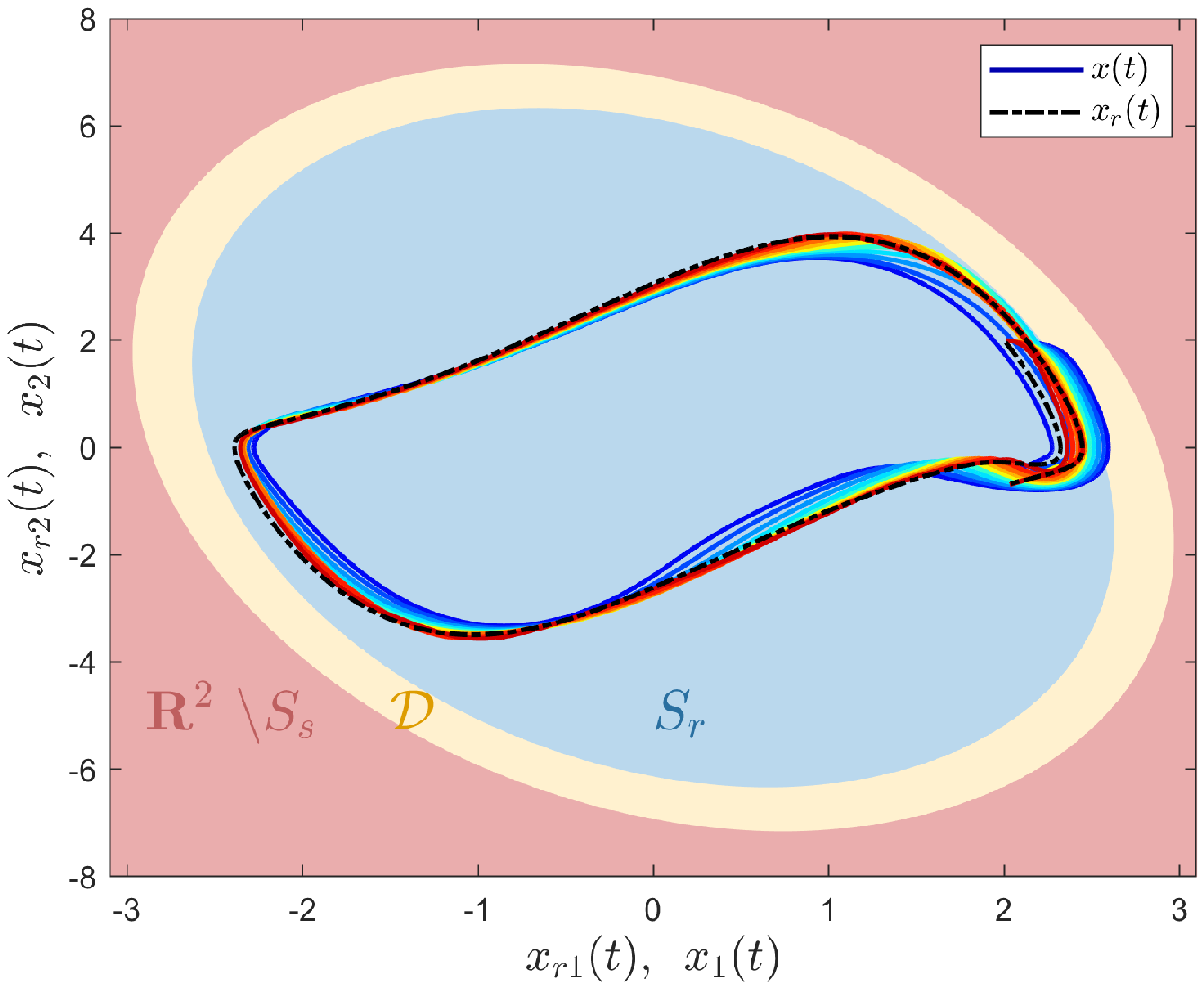} \vspace{-0.3cm}
	\caption{System state phase portrait with the proposed adaptive control architecture  with constant performance bound for $\gamma \in [0.05, 5]$ (blue to red).} \vspace{-.3cm}
	\label{fig8} \end{figure}

We now consider that only a set of reference system trajectories $S_r$ is known, where the safety margin can be characterized as a constant set $\mathcal{D}$. 
Based on the selected reference model, the reference set $S_r$ is selected as $S_r = \{x_\rom{r}(t): x_\rom{r}\mT(t) P x_\rom{r}(t) < 2.8\}$, where in this case the constant performance bound is selected as  $\bar \epsilon = 1.3$ to ensure safety.
We now apply the adaptive control signal in \eqref{uad} with the proposed update law in \eqref{Wdot} with different values of the adaptation rate $\gamma \in [0.05, 5]$. 
Figure \ref{fig8} shows that the proposed controller ensures safety of the system state trajectories where the obtained performance guarantee is independent of the selection of the adaptation rate $\gamma$.
This is also clear from  Figure \ref{fig9tv} where $h(e(t))$ is always positive resulting in $e(t) \in \mathcal{D} \ \forall t\geq0$.


Finally, Figures \ref{fig10} and \ref{fig11} compare the tracking performance for the proposed control architecture with constant and time-varying  performance bounds.
As expected, although a constant performance bound results in closer tracking of the reference trajectories, the control input is larger than the case with time-varying performance bound. 
In addition, one can see from Figure \ref{fig11} that the time-varying performance bound results in lower effective adaptation rate (i.e., $\gamma ({h(\cdot)+V(\cdot)})/{h^2(\cdot)}$ in \eqref{Wdot}), improving the robustness of the system. 

	
\begin{figure}[b!] \center \vspace{-.2cm}
	\includegraphics[width=8cm]{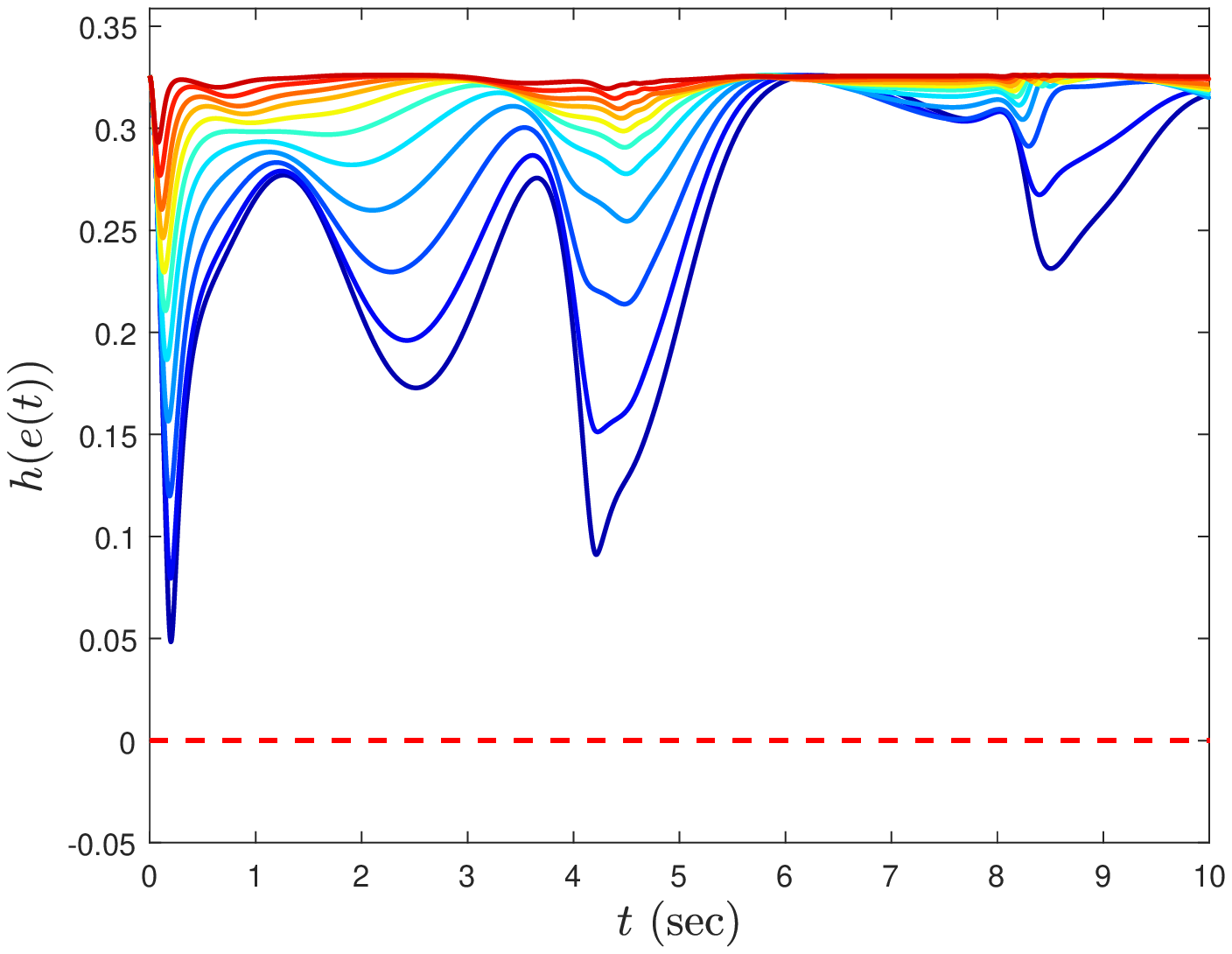} \vspace{-0.27cm}
	\caption{The evolution of $h(e(t))$ with the proposed adaptive control architecture  with constant performance bound for $\gamma \in [0.05, 5]$ (blue to red).} \vspace{-.25cm}
	\label{fig9} \end{figure}	
\begin{figure}[b!] \center \vspace{.0cm}
	\includegraphics[clip,trim=.2cm 0 .8cm .7cm, width=8.5cm]{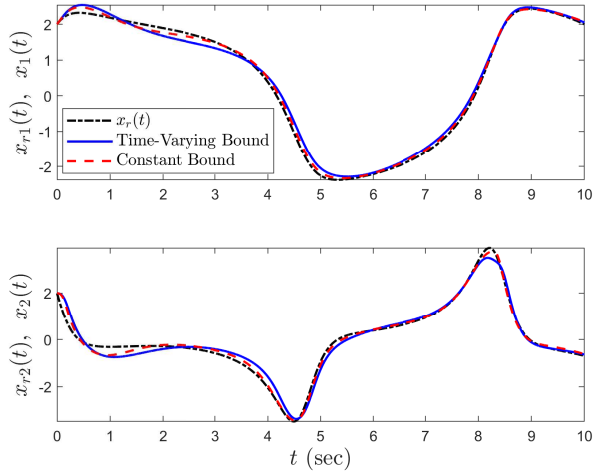} \vspace{-0.27cm}
	\caption{Comparison of the tracking performance using the proposed adaptive control architecture with  constant and  time-varying  performance bounds.} \vspace{-.25cm}
	\label{fig10} \end{figure}

\begin{figure}[b!] \center \vspace{.0cm}
	\includegraphics[clip,trim=.2cm 0 .8cm .7cm, width=8.5cm]{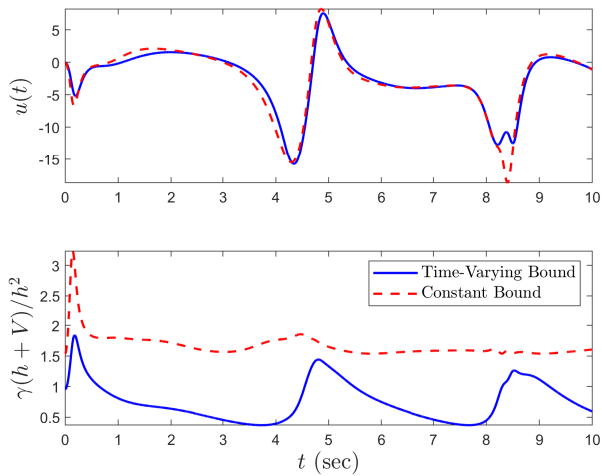} \vspace{-0.27cm}
	\caption{Comparison of the control signal (top) and effective adaptation rate (bottom) using the proposed adaptive control architecture with constant and time-varying  performance bounds.} \vspace{-.0cm}
	\label{fig11} \end{figure}

\section{Conclusion}
In this paper, we developed a new model reference adaptive control architecture based on nonlinear reference models for uncertain nonlinear systems.
Specifically, the key feature of the proposed approach was to suppress the effects of system uncertainties regardless of their magnitude and rate upper bounds. As a result, the system trajectories evolve within a user-specified prescribed distance from the nonlinear reference trajectories. Based on the safety specifications for a given system, this user-specified distance can be characterized to render the closed-loop system trajectories within the safe set, without the requirement of an ad-hoc tuning process for the adaptation rate.
An illustrative numerical example were further provided to demonstrate the efficacy of the proposed approach.


\bibliographystyle{IEEEtran} \baselineskip 12pt
{\footnotesize \bibliography{references}}
\end{document}